\documentclass{sigchi-ext}
\usepackage[T1]{fontenc}
\usepackage{textcomp}
\usepackage[scaled=.92]{helvet} 
\usepackage{graphicx} 
\usepackage{balance}  
\usepackage{booktabs} 
\usepackage{ccicons}  
\usepackage{ragged2e} 

\def\plaintitle{AI-Mediated Exchange Theory}

\def\emptyauthor{}
\def\plainkeywords{AI-mediated exchange theory, human-AI interaction, human-computer interaction, AI, algorithms, social exchange theory.}

\title{AI-Mediated Exchange Theory}

\numberofauthors{2}
\author{%
  \alignauthor{%
    \textbf{Xiao Ma}\\
    \affaddr{Cornell Tech, Cornell University} \\
    \affaddr{New York, NY 10044, USA} \\
    \email{xm75@cornell.edu} 
  }
  \vfil
  \alignauthor{%
    \textbf{Taylor W. Brown}\\
    \affaddr{Duke University} \\
    \affaddr{Durham, NC 27708, USA} \\
    \email{taylor.w.brown@duke.edu}
  }
}

\definecolor{linkColor}{RGB}{6,125,233}
\hypersetup{%
  pdftitle={\plaintitle},
  pdfauthor={\emptyauthor},
  pdfkeywords={\plainkeywords},
  bookmarksnumbered,
  pdfstartview={FitH},
  colorlinks,
  citecolor=black,
  filecolor=black,
  linkcolor=black,
  urlcolor=linkColor,
  breaklinks=true,
}

\begin{document}

\CopyrightYear{2020}
\setcopyright{rightsretained}
\conferenceinfo{CHI'20,}{April  25--30, 2020, Honolulu, HI, USA}
\isbn{978-1-4503-6819-3/20/04}
\doi{}
\copyrightinfo{\acmcopyright}

\maketitle

\RaggedRight{} 

\begin{abstract}

As Artificial Intelligence (AI) plays an ever-expanding role in sociotechnical systems, it is important to articulate the relationships between humans and AI.
However, the scholarly communities studying human-AI relationships --- including but not limited to social computing, machine learning, science and technology studies, and other social sciences --- are divided by the perspectives that define them.
These perspectives vary both by their focus on humans or AI, and in the micro/macro lenses through which they approach subjects.
These differences inhibit the integration of findings, and thus impede science and interdisciplinarity.
In this position paper, we propose the development of a framework \emph{AI-Mediated Exchange Theory} (AI-MET) to bridge these divides.
As an extension to Social Exchange Theory (SET) in the social sciences, AI-MET views AI as influencing human-to-human relationships via a taxonomy of \emph{mediation mechanisms}. We list initial ideas of these mechanisms, and show how AI-MET can be used to help human-AI research communities speak to one another.

\end{abstract}




\section{Introduction}
Relationships between humans and Artificial Intelligence (AI) are receiving increased attention across different research communities, driven by wide deployment of AI systems in the real world and increased awareness of the potential issues such deployment could bring.
Although there are many different ways of defining AI, here we use AI to refer to ``the designing and building of intelligent agents that receive percepts from the environment and take actions that affect that environment'' ~\cite{russell2016artificial}.
The human-AI research communities include but are not limited to: social computing (e.g., CHI, CSCW), Science and Technology Studies (STS), machine learning (e.g., fairness in machine learning~\cite{selbst2019fairness}), and the social sciences --- especially the emerging Computational Social Science (CSS).
Due to the interdisciplinary nature of the human-AI problem, communication across these research communities is necessary to drive scientific advancements that benefit both humans and AI.

Although often overlapping in topics of study (e.g., algorithmic ranking in social media news feeds), these research communities sometimes hold different perspectives on how human and AI systems are organized, which creates challenges when attempts are made to communicate or integrate findings.
We can view the different perspectives of these communities as spread along two axes: (1) humans/AI and (2) micro/macro.
The first axis describes whether the research focuses more on understanding or designing for humans in sociotechnical systems, or more on building and improving the AI systems involved.
The second axis describes the level of abstraction of humans --- research at the micro end focuses more on individual interactions (e.g., how an individual interacts with the algorithms in social media feeds), while research at the macro end of the axis focuses on larger social structural processes (e.g., how do algorithms affect markets, healthcare, or the concept of justice in law?).

Although in many cases research perspectives don't lie cleanly at the ends of the axes, we can often find example ``pairs'' of projects from different research communities that spread out along the poles, even if they address the same topic.
For instance, on the human-side of the first axis, Eslami et al. presents a qualitative interview study on how users perceive Facebook news feed ranking~\cite{eslami2015always}.
By contrast, most work on recommendation systems lies closer to the AI-end of the axis, focusing on the development of algorithms using large-scale datasets to optimize for specific algorithmically-relevant and statistically defined measures (e.g., Twitter news feed ranking that might optimize for engagement~\cite{phelan2009using}).
For the micro/macro axis, Kizilcec conducted a user study to show how transparency affects trust in algorithms in peer-to-peer evaluation settings, which focused on the micro behavior of individuals~\cite{kizilcec2016much}.
By contrast, the discussion around regulation on algorithmic transparency of algorithms is often more macro-level, from the perspective of philosophy and the social sciences~\cite{mittelstadt2019explaining,miller2018explanation}.

The spread of different research perspectives on the humans/AI and micro/macro axes creates gaps in the inter-community learning and collaboration.
It is not immediately obvious how research lying at different coordinates can learn from each other.
The topography of the different research perspective spread along the humans/AI and micro/macro axes entails gaps that trouble their ability to communicate and collaborate, despite the common call for interdisciplinary approaches~\cite{selbst2019fairness,boyd2011six}.
In an ideal world, practitioners building AI systems would bear human factors, including user behavior and ethics, in mind.
They would consider the risks and the benefits of an algorithm on people's lives.
Likewise, scholars of modern social systems would consider the role of AI --- especially when data are gathered from an online platform in which AI systems affect how people behave.
Such an ideal world would also see micro-level research, such as specific studies in HCI, effectively informing macro-level research, such as law and policy.
Falling short of such research integrations holds the potential for serious societal consequences~\cite{barocas2016big}, while successfully doing so may lead to benefits for both science and society~\cite{green2019disparate}.

\section{AI-Mediated Exchange Theory}
For the reasons listed above, there is tremendous value in creating a theoretical framework that applies equally across the humans/AI and micro/macro axes.
This position paper proposes the development of \emph{AI-Mediated Exchange (AI-MET)} as that framework.
As an extension to Social Exchange Theory (SET) --- a well-established theory in the social sciences --- AI-MET is not an attempt to develop yet another area of research.
Rather, it is proposed as a framework through which the several existing communities of research on the topic of human-AI interaction can speak to one other.

Traditionally, SET envisions society as a configuration of exchange relationships between actors, be they individuals or institutions~\cite{cook_two_1992}.
There are two types of exchange in SET, \emph{direct} and \emph{generalized}, which we go into in more detail throughout the extended version of our paper.
Together, the direct and generalized forms of exchange link micro-level social interactions to macro-level concepts such as power, communal resilience, and trust~\cite{cook2005trust}.
For our work, we reap this benefit of micro/macro bridging in SET, and extend it by inserting AI as a mediator of human social exchange relationships, thereby also bridging the humans/AI divide previously discussed.

The core of the proposed AI-MET framework is a taxonomy of AI mediation mechanisms, which we derived from a qualitative coding of human-AI literature.
We show that this taxonomy can be used to clarify the role of AI in mediating social exchanges.
We then apply AI-MET to concrete examples (social media feed ranking as an example of generalized exchange, and algorithmic hiring as an example of direct exchange) to demonstrate the utility of AI-MET as a means integrating human-AI research communities.

\begin{figure}[h!]
  \centering
  \includegraphics[width=0.8\linewidth]{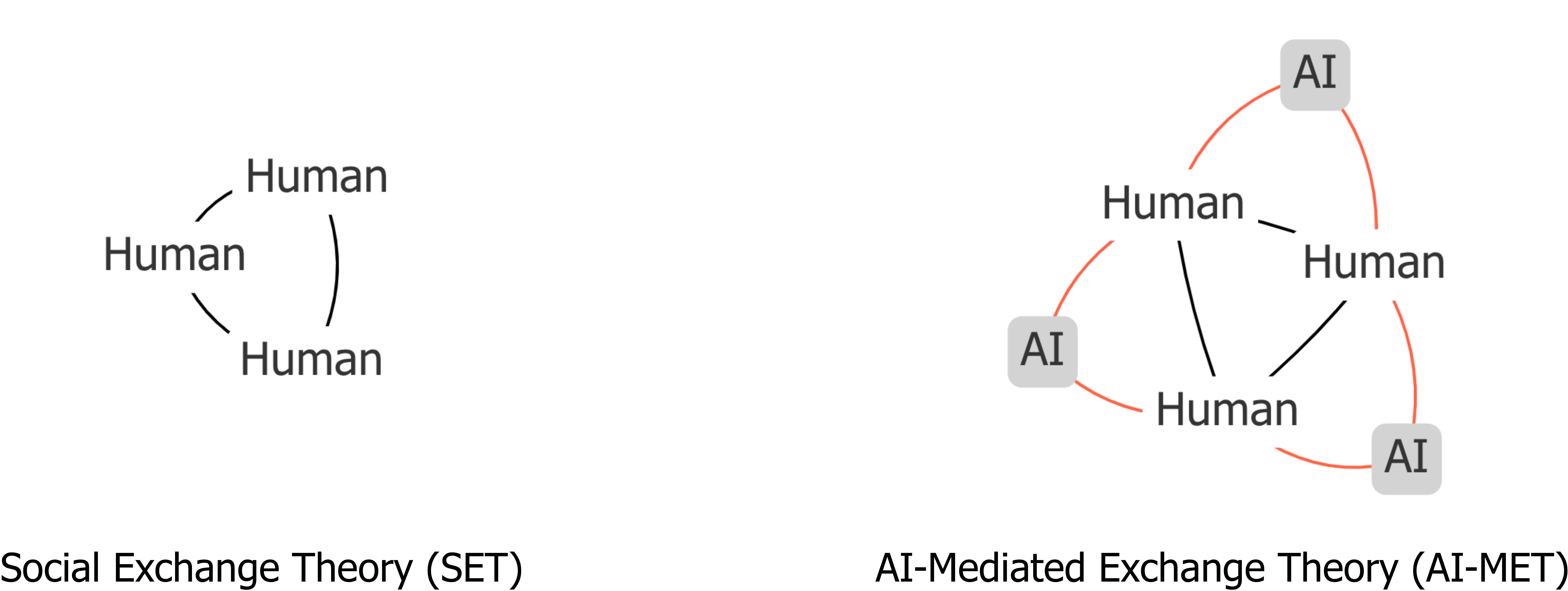}
  \caption{Theoretical illustration of Social Exchange Theory (left) and AI-Mediated Exchange Theory (right)}
  \label{fig:ai-met}
\end{figure}

The theoretical setup of the AI-Mediated Exchange Theory is conceptually simple.
Social Exchange Theory states that the world can be viewed as a series of exchange relationships, as illustrated on the left side of ~\autoref{fig:ai-met}.
Each node represents a social actor, and edges represent a specific type of exchange/interaction.
These exchanges could be financial, informational, or simply relational, such as friendships.
Three nodes represent the minimal setup for ``group generalized exchange'', where people pool resources together to produce greater value, and then redistribute the resources at a later time, such as in communities and small social groups.
AI-Mediated Exchange Theory, depicted on the right side of~\autoref{fig:ai-met}, shows how AI systems can be incorporated into these social exchange systems by mediating social exchange relationships.
The nodes ``AI'' here represents an autonomous system that can act with agency and affect the environment it is in~\cite{russell2016artificial}.

In conclusion, this position paper sets up the motivation for extending Social Exchange Theory (SET) into AI-Mediated Exchange Theory (AI-MET), which we will fully develop in an extended paper.
With the full theoretical framework and concrete mediation mechanisms  derived from qualitative work (e.g., AI curation and AI matching), AI-Mediated Exchange Theory can help explain how AI mediates social exchange in sociotechnical systems, as well as allow various human-AI research communities to integrate.

\balance{} 

\bibliographystyle{SIGCHI-Reference-Format}
\bibliography{main}

\end{document}